\documentclass[11pt]{article}
\usepackage[textwidth=15.2cm,textheight=22cm]{geometry}
\usepackage{amsmath,amssymb}
\usepackage{latexsym}
\usepackage{multicol}
\usepackage{color}
\usepackage{graphicx}
\usepackage{tikz}
\usetikzlibrary{shapes,backgrounds}

\usepackage{bm}
\allowdisplaybreaks[1]

\newcommand{\be}{\begin{equation}}
\newcommand{\ee}{\end{equation}}
\newcommand{\ba}{\begin{eqnarray}}
\newcommand{\ea}{\end{eqnarray}}
\newcommand{\bdm}{\begin{displaymath}}
\newcommand{\edm}{\end{displaymath}}

\newcommand{\rom}[1]{\uppercase\expandafter{\romannumeral #1\relax}}

\def\ba{\bar A}

\def\beq{\begin{equation}}
\def\eeq{\end{equation}}

\def\bea{\begin{align}}
\def\eea{\end{align}}
\def\beas{\begin{eqnarray*}}
\def\eeas{\end{eqnarray*}}
\def\sla{\raise.15ex\hbox{$/$}\kern-.57em}

\def\spa#1.#2{\left\langle#1\,#2\right\rangle}
\def\spb#1.#2{\left[#1\,#2\right]}

\begin{document}
\begin{titlepage}
\vskip 1cm
\centerline{\LARGE{\bf {Hadronic Strings}}}
\vskip .5cm
\centerline{\LARGE{\bf { - }}}
\vskip .5cm
\centerline{\LARGE{\bf { a Revisit in the Shade of Moonshine}}}

\vskip .5cm
\centerline{\bf Lars Brink }
\vskip .5cm
\centerline{\em Department of Physics}
\centerline{\em Chalmers University
of Technology, }
\centerline{\em S-412 96 G\"oteborg, Sweden}
\vskip .5cm
\centerline{\em and}
\vskip .5cm
\centerline{\em Arnold Sommerfeld Center for Theoretical Physics}
\centerline{\em Ludwig Maximilian University of Munich}
\centerline{\em Theresienstr. 37, D-80333 M\"unchen, Germany}

\vskip 1.5cm

\centerline{\bf {Abstract}}
\vskip .5cm
\noindent When asked to write a contribution to the memorial volume for Peter Freund I went through my memory of the first time I met Peter. This was in 1971 when Dual Models were very popular and I had just joined in the efforts. For a number of years I worked on the problem of finding a realistic Dual Model/String Theory for hadrons, and here I will review those efforts as they happened, but also in the light of what we now know about hadrons from QCD. I will argue for when a string picture of hadrons is appropriate and discuss its limitations and the specific results you get from it. 
\vskip 3.5cm
{\hskip 5cm
"{\it Till flydda tider {\aa}terg{\aa}r min tanke \"an s{\aa}  g\"arna"}}

\hskip 10.5cm J. L. Runeberg

\vskip 2cm

\it{Contribution to the Memorial Volume for Peter Freund}

\end{titlepage}

\section{Encounters with Peter}
I came to CERN in June of 1971 as a fellow. I came a few months earlier than the other fellows since the Theory Department wanted to lift me out from my home institution and give me an early chance. i did not even have a Ph.D at the time. I went to all seminars that were daily at the time and one of the first ones was given by Peter Freund. He had spent a year at Imperial College and he talked about diffraction models using duality~\cite{Freund:1971tw}, but he did not mention any dual models as far as I remember. Already then I saw that he was a forceful person having grown up in post-war Timisoara and hardened by the student revolt there in 1956. 

Being in Chicago close to Yoichiro Nambu he was of course involved in Dual Models, and he was certainly one of those whose papers we studied carefully at CERN at the time. He was also early to think about putting quarks on the string~\cite{Freund:1972gw}, a subject I will come back to. Our interests merged again later in the 70's when Peter got interested in supersymmetry and then in supergravity. His paper from 1980 with Rubin~\cite{Freund:1980xh} certainly shook up the supergravity community. 

Our paths met again when he got into string theory later in the 80's, and again we took great notice when he introduced p-adic numbers~\cite{Freund:1987ks} and adelic string amplitudes~\cite{Freund:1987ck}, concepts that I had not even heard about before. Many papers in his later career involved intriguing mathematics with his deep knowledge in modern particle physics as well as in more formal theory. Others will describe better his scientific achievements and his influence on modern fundamental physics. Peter was a strong member of the  Enrico Fermi Institute in Chicago all through his long life. He was one of the few who came really close to Nambu and saw his extraordinary talent for physics, and he was a strong lobbyist for the Nobel Prize that Nambu eventually got.  

Peter was the chairman of the Schr\"odinger Society and was the remaining link to Erwin Schr\"odinger, whom he had met when he was a graduate student in Vienna. I had the privilege to give the Schr\"odinger lecture some ten years ago, and at the same time he gave me a long sight-seeing of Timisoara which had then just become popular by the fascinating and tragic descriptions that Hertha M\"uller had given. She was a favourite of both of us. He then showed that he was a renaissance man being home both in Europe and the USA, in the old quarters of the once so cultivated Central Europe as well as in the most modern parts of the great cities in the USA. A few months before he died he sent me an email where he described his disease and told me that he only had a few months more to live. It was a shock. He was such a strong man both physically and mentally and was not supposed to leave us so early.

\section{Dual Models/String Theory: The first few years}
After the bomb shell created by Gabriele Veneziano in 1968 when he constructed a crossing - symmetric, Regge behaved amplitude for linearly rising trajectories for the scattering amplitude $\pi\,\,\pi \rightarrow \pi\,\,\omega$  \cite{Veneziano:1968yb}, the Veneziano Model developed quite quickly with an average of one paper a day for the next years. The model described $\pi\, \pi$-scattering and soon $N-$point amplitudes were constructed~\cite{Chan:1969ex}, \cite{Koba:1969rw}, a string picture emerged~\cite{Nambu:1969se}, \cite{Susskind:1970xm}, \cite{holger}, and the first crack in the hadronic picture appeared when Viraosoro showed~\cite{Virasoro:1969zu} 
 that only for the model with a massless spin-one particle does it have an infinite symmetry that could provide a ghost-free spectrum.
 
 Another problem that was solved was whether the amplitudes could factorize. Indeed they could as Nambu~\cite{Nambu:1969se} and Fubini, Gordon and Veneziano~\cite{Fubini:1969wp} discovered. They found that the amplitudes could be written as
\be \label{factor}
B_N = <0| V\,D\,V\,......D\,V|0>,
\ee
with $D$ a propagator and $V$ a vertex operator.

This opened the way to study the spectrum using the Virasoro algebra, Much have been written about the early history so I have been and will be extremely brief and selective in my account of the early history here. I only do this in order to set the stage for what I will describe in more detail. There is a tendency in modern Superstring Theory to leave out this period that was instrumental for the development of the theory.

\subsection{1971-73}
The first two years of these three years were exceptional with one discovery coming after the other. It started already in the last days of 1970 when Pierre Ramond~\cite{Ramond:1971gb} out of nowhere found the spectrum and the symmetries of the fermionic string. He did it by following Dirac's original analysis extending not only the spacetime coordinates but also the $\gamma$-matrices to depend on a world sheet variable. This was not only the first piece in the Superstring Theory but also in supersymmetry.

 Shortly after Andr\' e Neveu and John Schwarz~\cite{Neveu:1971rx} discovered a new bosonic string theory using two-dimensional supersymmetry and Charles Thorn~\cite{Thorn:1971jc} a model with two fermions and an arbitrary number of bosons, which was to become the 'Ramond model/sector'. Although the models contained unphysical elements such as tachyons and unwanted fermion states, the embryo of the Superstring Theory was born. An important addition came when Edward Corrigan and David Olive~\cite{Corrigan:1972tg} constructed the fermion emission vertex making it possible to construct amplitudes with more than two fermions. The final merging of the two sectors came when I together with Olive, Claudio Rebbi and Jo\"el Scherk~\cite{Brink:1973jd} proved that the two sectors are connected in a unitary way through the fermion emission vertex satisfying the gauge conditions on both sides. This was possible to prove with the help of the physical state projection operator that I had constructed with Olive~\cite{Brink:1973qm}. Some years later the unphysical states in the two sectors were removed by Ferdinando Gliozzi, Olive and Scherk~\cite{Gliozzi:1976jf} and the Superstring Theory was ready to fly even though it took some years for it to do so.
 
 Another important ingredient was the discovery of Claude Lovelace~\cite{Lovelace:1971fa} that the so-called twisted loop in the Veneziano model could only be consistent in $d=26$. A twisted loop in a string picture is a loop diagram, where the incoming strings are attached at one end of the string and the outgoing at the other.  Since the string is conformally invariant you can stretch the diagram such that the external states become normal states. Then the diagram is just the scattering of two open strings with an intermediate state that is a closed string, ie. the loop diagram have been turned into a tree diagram with a new set of intermediate states, the states of a closed string. The trajectory of it was called the Pomeron trajectory even though it had intercept 2.
 
 The critical dimension of $26$ was not taken seriously until Richard Brower~\cite{Brower:1972wj} and Peter Goddard and Charles Thorn~\cite{Goddard:1972ky} proved that in the critical dimension the Veneziano model has a positive-norm spectrum. Similarly they found that in the Ramond-Neveu-Schwarz model the critical dimension for a positive-norm spectrum is $10$. This was further clinched when Goddard, Rebbi and Thorn joined up with Jeffrey Goldstone~\cite{Goddard:1973qh} to study the quantisation of the relativistic string described by an action invented by Nambu~\cite{NG} and Tetsuo Got\={o}, \cite{Goto:1971ce}.
 
 \section{Self-consistency conditions for string theories}
 Since we were all at the time searching for a dual model/string theory that could describe meson scattering, the dream at the time was to find a realistic model extended from the existing models. The idea was that if such a model could be found, it should have critical dimension $4$ and have a positive norm spectrum consisting of the known meson trajectories. This was before QCD and we only knew of three quarks. Holger Bech Nielsen and I then searched for a general scheme which all possible string theories could satisfy. 
 
 We knew that the key to understand the critical dimension, originating in the work of Lovelace, was the remarkable properties of the partition function that entered into the loop calculation. These arise from the Jacobi $\theta$-functions described by Jacobi in the 1830's together with his assistant J. Scherk. Schematically they relate the various $\theta$-functions when one of its variable, we can call it $\tau \rightarrow \frac{1}{\tau}$, 'the Jacobi imaginary transformation'. (We have made it real.)
 
 Consider now the calculation by Lovelace stripped by the external states.
 
 \begin{center}
	\begin{tikzpicture}
	
	\node[thick, circle, minimum size=1.6cm, draw, fill=white] (a) {};
	\node[thick, circle, minimum size=0.8cm, draw, fill=white] (b) {};
	\draw[black, thick] (1.7,0) -- (2,0);
	\draw[black, thick] (1.7,0.1) -- (2,0.1);
	\node (c) [thick, xshift= 4cm, cylinder, shape border rotate = 180, draw, minimum height=1.6cm, minimum width=1.4cm] {};
	\end{tikzpicture}
\end{center}
\begin{center}
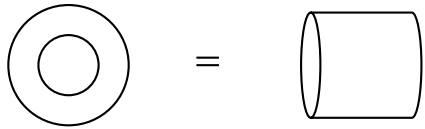
Fig.1 {\it Duality property of an open string loop}
\end{center}
 
 When these calculations are performed we first have made a Wick rotation of the two-dimensional worldsheet. We can fix the gauge such that the remaining symmetry is conformal symmetry. This means that we can stretch the worldsheet diagram such that the loop of the open string has become a propagating closed string. The open string satisfies Neumann boundary conditions. For the closed string propagator this amounts to the propagation of a string whose initial and final state have momentum density zero, a vacuum string. The calculations is not straightforward to do and it will be highly divergent. Also at the time  of these considerations it was not known how to consistently compute a one-loop diagram. This was only solved in 1973 by Olive and myself~\cite{Brink:1973gi} using Feynman's tree theorem and our physical state projection operator~\cite{Brink:1973qm}. Also since the idea was to consider strings with quarks at the ends this was not the ideal scheme, since  this would only have led us to expressions with couplings of quarks to pomerons.
 
 Instead we considered the following diagram, Fig.2, the propagation of a 'vacuum string'~\cite{Brink:1973kn}.
 
  
 \begin{center}
 	\begin{tikzpicture}
 	 \draw[black, thick] (2,0) -- (2,1.5)node[midway, xshift= -1.1cm]{$|P(\sigma)=0 \rangle$};
 	\draw[black, thick] (2,0) -- (4.5,0);
 	\draw[black, thick] (4.5,0) -- (4.5,1.5)node[midway, xshift= 1.1cm]{$|P(\sigma)=0 \rangle$};
 	\draw[black, thick] (2,1.5) -- (4.5,1.5);
 		\draw[black, thick] (7.5,0.6) -- (7.8,0.6);
 	\draw[black, thick] (7.5,0.7) -- (7.8,0.7);
 	\draw[black, thick] (9,0) -- (9,1.5);
 	\draw[black, thick] (9,0) -- (11.5,0)node[midway, yshift=-0.6cm]{$|P(\sigma)=0 \rangle$};
 	\draw[black, thick] (11.5,0) -- (11.5,1.5);
 	\draw[black, thick] (9,1.5) -- (11.5,1.5)node[midway, yshift= 0.6cm]{$|P(\sigma)=0 \rangle$};
 	\end{tikzpicture}
 \end{center}
 \begin{center}
 Fig. 2 {\it Propagation of a 'vacuum string'}
  \end{center}
 
 At the time there were several useful ways to compute this diagram. One was the analogue model of Fairlie and Nielsen~\cite{Fairlie:1970tc}, where the problem is mapped to the calculation of electric currents on a conducting sheet, another Nielsen's formulation~\cite{Nielsen:1969} in terms of infinitely dense Feynman diagrams, both precursors of the functional integration over the string action.
 
 In a functional formulation this can formally be written as being proportional to
 
 \be \label{2}
 \int d\tau \int \textit D\,{ \Phi\,\, \psi^{(f)}}^\dagger {}_{p_\mu}(\sigma_1, \sigma_2=\sigma_2{}^f)\, e^{-\int \, \textit L (\Phi)\, d^2 \sigma}\,\psi^{(i)}{}_{p_\mu}(\sigma_1, \sigma_2=0),
\ee
where we integrate over all independent string fields ($\Phi$). Again we have made the assumption that we have gauge fixed to a conformal gauge and work in a Euclidean world sheet. We restrict ourselves to simply connected surfaces. Since the initial and final wave functions $\psi^{(i)}{}_{p_\mu}$ and $\psi^{(f)}{}_{p_\mu}$through their end points each define two
points on the boundary of the surface, a natural choice for the surfaces is to choose
rectangles with the end points of the strings at the corners. The conformally non-equivalent ways of attaching the end points can then be parametrized by the ratio
of the lengths at the sides of the rectangle, $\tau$ stands for a function of this ratio chosen such that the integration over $\tau$ gives the full amplitude.

We now further assume that the underlying string has a bosonic part of conventional type, ie. coming from the spacetime dependence of the string. This means that the propagator has the conventional form in terms of the generator $L_0$ of the Virasoro algebra. This also means that we allow for strings with fermionic degrees of freedom and other bosonic degrees as well as insertions of quarks at the ends of the string. We can then write the integrand in (\ref{2}) as

\begin{align}\label{3}
\nonumber &   \int \textit D\, {\Phi\,\, \psi^{(f)}}^\dagger {}_{p_\mu}(\sigma_1, \sigma_2=\sigma_2{}^f)\, e^{-\int \, \textit L (\Phi)\, d^2 \sigma}\,\psi^{(i)}{}_{p_\mu}(\sigma_1, \sigma_2=0) \\
 &= \displaystyle \sum_n < \psi^{(f)} | n> \, e^{R_1(\tau)(p^2-m_n^2)} \,<n |\psi^{(i)}>.
 \end{align}
 
 Here $R_1(\tau)$ is what distinguishes the different conformal diagrams. In a diagram with sides a and b
 
 \be
 R_1 = \frac{a}{b} \pi \alpha^\prime,
 \ee
 where $\alpha^\prime$ is the Regge slope of the particle trajectories which we will put equal to $1$ from now on. In an analogue model $R$ is essentially the resistance for the flow.
 
 If we inspect Fig. 2 we see that we can equally well compute the diagram in the vertical directions. The Neumann boundary conditions we have as the string propagates horizontally means vertically that the initially and final state again is a vacuum string. The vacuum string in the horizontal directions means Neumann boundary conditions in the vertical direction. Remember that we are in a Euclidean world sheet. We now specify the states  $|\psi^{(i)}>$ and $|\psi^{(f)}>$ to be the vacuum string and specify to $ p^\mu(\sigma) =0$. We can then perform the calculation (\ref{3}) in two ways and must get an identity
 
 \be \label{5}
 \int d\tau (R_1)  \displaystyle \sum_n |< vac| n>|^2 \, e^{-R_1 m_n^2} = \int d\tau (R_2)  \displaystyle \sum_n |< vac| n>|^2 \, e^{-R_2 m_n^2},
\ee
where we have allowed for a more general dependence on $R_i$. In the Veneziano model and the Neveu-Schwarz model $Ê\tau = R_i$. 
The two sides are related by 
\be \label{6}
R_1\, R_2 = \pi^2,
\ee
which means that $dR_1R_2 + R_1 dR_2 = 0$ and $dR_2 = - R_2/R_1 dR_1$.

We can now check this in the bosonic string. For $| n>$ we take
\be
| n> = \Pi \, \frac{{(a^{\dagger}{}^i_n)}^{\lambda^i_n}}{\sqrt {\lambda^i_n}!}\,|0>,
\ee
where the harmonic oscillators can be taken to be either the transverse ones after a choice of the light-cone gauge or the DDF-operators~\cite{DelGiudice:1971yjh}.

We can solve first the one-dimensional case. The vacuum string is defined by\footnote{This is not a well-defined state, which we should keep in mind.}
\be
a^\dagger\, |vac> = a \,|vac>.
\ee
 Then 
 \begin{align} \label{9}
  \nonumber &|< vac\,|\frac{{a^\dagger}^n} {\sqrt n!}|0> |^2 =  |< vac\,|\frac{a{a^\dagger}^{n-1}} {\sqrt n!}|0> |^2\\
 \nonumber  &= \frac{(n-1)}{n!}\, <vac\,| {a^\dagger}^{n-2}|0> = .....\\
&=\frac{(n-1)!!}{n!!} |<vac\,|0>|^2.
 \end{align}
This is for even $n$, for odd $n$ one gets zero. We can now solve the corresponding sum.
\begin{align}\label{10}
\nonumber &\displaystyle \sum_{n=0} ^\infty\, \frac{(n-1)!!}{n!!} \, e^{-nR} \\
 =&\displaystyle \sum_{k=0 }^\infty\,(-1)^k\binom{-1/2}{k}\, e^{-2kR}= (1-e^{-2R})^{-1/2}.
\end{align}
 
 Knowing that the lowest state has $m^2 = -1$, we can now extract $e^{R}$ from the sum in (\ref{5}) and use the differentiation of (\ref{6}) and extract the integrands in (\ref{5}), taking care of the integration limits and finally use the results in (\ref{9}) and (\ref{10}) to get the equation 
 \be
 R_1\, e^{R_1} \, \displaystyle \prod_r\,(1-e^{-2R_1 r})^{-D/2}|<vac|0>|^2 =R_2\, e^{R_2} \, \displaystyle \prod_r\,(1-e^{-2R_2 r})^{-D/2}|<vac|0>|^2,
\ee
where $D$ is the number of transverse dimensions. Dividing out the overlap $|<vac|0>|^2$~\footnote{This expression is again not well-defined but we can use some regularisation to make the calculation consistent.},  choosing $D=24$ 
\be
 {R_1}\,  e^{R_1} \, \displaystyle \prod_r\,(1-e^{-2R_1 r})^{-12} =R_2\, e^{R_2} \, \displaystyle \prod_r\,(1-e^{-2R_2 r})^{-12}.
\ee
We can compare this to the Jacobi imaginary transformation on $\theta_1({0,\tau})$.
\be
 {R_1}^{-1/4} e^{R_1/12} \, \displaystyle \prod_r\,(1-e^{-2R_1 r})^{-1} =R_2^{-1/4} e^{R_1/12} \, \displaystyle \prod_r\,(1-e^{-2R_2 r})^{-1}.
\ee

We find that all exponential terms match. It is only the power term that does not match. When we did this calculation in 1972 we had no explanation for it. It was only with the advent of the BRST formulation of the string that it could be understood~\cite{Misha}. The point is that the vacuum state we have used is not fully a physical state. In order to find such a one we have to introduce ghost insertions at the corners. However, the discrepancies are not preventing us from finding the possible string models. The only problem for the vacuum string to be physical comes from the corners of the diagram. To correct for that can never effect the infinite series of exponential terms. Hence we can confidently use this method to look for new strings models. We~\cite{HBNLB2} also performed the calculation above for the Neveu-Schwarz model and found that all exponential terms match for $10$ spacetime dimensions and the correct intercept.

Some years later Werner Nahm~\cite{Nahm:1976dj} made a study of possible partition functions and could also describe the spectrum of the heterotic string. He did not formulate an interacting model, and the idea to construct such a one came almost a decade later~\cite{Gross:1984dd}. Our next step after having set up the scheme with the 'vacuum string' was not to construct more models of the "conventional" type but to look for realistic models for hadron scattering. Setting up the calculations as in Fig. 2 could lead to a full classification of the type Nahm performed.

 \section{Self-consistency conditions for string theories with $SU(6)$ quantum numbers}
 
 After finding that our self-consistency equations work for the at the time known  string theories/dual models, our next step was to consider strings with real meson quantum numbers. At the time only three quarks were known and we assumed that in a  correct dual model for nature, if such a theory exists, the mesons can be constructed as strings with quarks attached to their ends. When we talk of quarks here we just mean something that carries quark quantum numbers. In 1972 we knew very little about quark masses and we assumed them to be zero for the calculation. I will here follow a paper which I wrote at the time together with Nielsen~\cite{Brink:1974cm} which we only published a few years later.

We now extend our considerations in sect. 3 to the propagation of a string with a quark ($a$) attached to one end, an antiquark ($\bar b$) attached to the other end, which develops into a string with an antiquark ($\bar c$) removed from the first end and a quark ($d$) removed from the second end.

 \begin{center}
	\begin{tikzpicture}
	\draw[black, thick] (2,0) -- (2,1.5)node[ xshift= -0.3cm]{$a$};
	\draw[black, thick] (2,0) -- (2,1.5)node[  xshift= -0.3cm,yshift= -1.5cm]{$\bar b$};
	\draw[black, thick] (2,0) -- (2,1.5)node[midway, xshift= -1.1cm]{$|P(\sigma)=0 \rangle$};
	\draw[black, thick] (2,0) -- (4.5,0)node[ xshift= 0.3cm]{$ d$};
	\draw[black, thick] (2,0) -- (4.5,0);
	\draw[black, thick] (4.5,0) -- (4.5,1.5)node[midway, xshift= 1.1cm]{$|P(\sigma)=0 \rangle$};
	\draw[black, thick] (4.5,0) -- (4.5,1.5)node[ xshift= 0.3cm]{$\bar c$};
	\draw[black, thick] (2,1.5) -- (4.5,1.5);
	\draw[black, thick] (7.5,0.6) -- (7.8,0.6);
	\draw[black, thick] (7.5,0.7) -- (7.8,0.7);
	\draw[black, thick] (9,0) -- (9,1.5);
	\draw[black, thick] (9,0) -- (9,1.5)node[ xshift= -0.3cm]{$a$};
	\draw[black, thick] (9,0) -- (9,1.5)node[  xshift= -0.3cm,yshift= -1.5cm]{$\bar b$};
	\draw[black, thick] (9,0) -- (11.5,0)node[midway, yshift=-0.6cm]{$|P(\sigma)=0 \rangle$};
	\draw[black, thick] (11.5,0) -- (11.5,1.5)node[ xshift= 0.3cm]{$ \bar c$};
	\draw[black, thick] (9,0) -- (11.5,0)node[ xshift= 0.3cm]{$ d$};
	\draw[black, thick] (11.5,0) -- (11.5,1.5);
	\draw[black, thick] (9,1.5) -- (11.5,1.5)node[midway, yshift= 0.6cm]{$|P(\sigma)=0 \rangle$};
	\end{tikzpicture}
\end{center}

\begin{center}
Fig.3 {\it Propagation of a 'vacuum' string with a pair of quark-antiquark at the ends}
\end{center}
As in the previous section the starting point is then

 \be \label{14}
 \int d\tau \int \textit D\,{ \Phi\,\, \psi^{(f)}}^\dagger {}_{p_\mu,\bar c \,d \,removed} \, e^{-\int \, \textit L (\Phi)\, d^2 \sigma}\,\psi^{(i)}{}_{p_\mu, a\, \bar b\, attached}.
\ee

We can follow through the considerations in sect. 3 up to the point when we specialize to the propagator of a string in the state  defined by being such that it couples to the physical vacuum. This time we consider a similar state with vacuum quantum numbers locally along the string except at the end points of the string where it has a quark at one end and an antiquark at the other end in the initial state and an antiquark and a quark removed from the ends in the final state. The state $|P(\sigma)=0 \rangle$ symbolises a state with vacuum quantum numbers in all its degrees of freedom, not only the momentum density. The total quantum numbers of the string state considered is thus those of a quark-antiquark pair, i.e. those of mesons. It may thus be assumed that these more general vacuum string states with quarks added or removed at the ends have significant overlaps with meson states, a property of which we shall make use below. Performing the same steps as in the previous section we are led to

\begin{align} \label{15}
\nonumber& \int d\tau \int \textit D\,{ \Phi\,\, \psi^{(f)}}^\dagger {}_{p_\mu,\bar c \,d \,removed} \, e^{-\int \, \textit L (\Phi)\, d^2 \sigma}\,\psi^{(i)}{}_{p_\mu, a\, \bar b\, attached} \\
& =   \int d\tau  \displaystyle \sum_n < vac, \bar c\,d\,removed\,| n>\,< n| vac,\,a\,\bar b\, attached\,> e^{-R m_n^2 }.
 \end{align}

Since we are working in a Euclidean space we now want to perform the same calculation with the roles of $\sigma_1$ and $\sigma_2$ interchanged, so that this time $\sigma_2$ is the development parameter while $\sigma_1$ is the parameter along the string. With quarks on the strings the self-consistency condition analogous to eq. (\ref 5) must express the equality between two ways of calculating essentially a $q\,\bar q$ scattering amplitude. More exactly the group theory follows the calculations of such an amplitude, while the string calculations are for a one-dimensional gluon string connecting the quarks (in modern terms).

If we write the amplitude as a propagator for a string in two different ways by having string wave functions of locally vacuum type attached to the two different pairs of parallel sides of the region $[0, \sigma_1^f] \times [0, \sigma_2{}^f]$ we obtain the self-consistency relation

\begin{align}\label{16}
\nonumber \int d\tau(R_2)    \displaystyle \sum_n < vac, \bar c\,d\,removed\,| n>\,< n| vac,\,a\,\bar b\, attached\,> e^{-R_2 m_n^2 } \\
= - \int d\tau(R_1)    \displaystyle \sum_n < vac, \bar b\,d\,removed\,| n>\,< n| vac,\,a\,\bar c\, attached\,> e^{-R_1 m_n^2 }.
\end{align}
Again the integration variables are connected through (\ref{6}). The minus sign in the relation is due to Fermi-Dirac statistics. Using (\ref{6}) we can again compare the integrands of the expression, and these are essentially the partition functions that should describe mesonic strings.

When we talk about quark quantum numbers we mean that a quark can be in six states, $u\uparrow, u \downarrow, d\uparrow, d \downarrow, s\uparrow, s\downarrow$, corresponding to the $SU(6)$ representation $\bf{6}$. The arrows indicate spin up or spin down. The symbols a, b, c and d can hence take on six values each. There are thus in principle $6^4$ relations (\ref{16}). Most of them, however, are trivial because of the conservation laws.

In order to confront the relations (\ref{16}) with experiments we assume $SU(6)$ invariance in the form
\begin{align}\label{17}
\nonumber& \displaystyle \sum_n < vac, \bar c\,d\,removed\,| n>\,< n| vac,\,a\,\bar b\, attached\,>\\
& = const < vac, \bar c\,d\,removed\,| vac,\,a\,\bar b\, attached\,>,
\end{align}
with the same constant for all quark combinations\footnote{I remind you again about the badly defined vacuum state, which needs a regularisation procedure.}. The sum is over all states belonging to the $\bf{35}$ and the corresponding singlet. We have hence added to the assumption about $SU(6)$ invariance the assumption that the singlet couples in the same way as the $\bf{35}$.

This $SU(6)$ assumption is like assuming $SU(6)$ invariance for coupling constants but not for the masses, since a matrix element $| vac,\,a\,\bar b\, attached\,>$ can be considered as a coupling of a quark-antiquark pair to a meson.

In the case of mesons with the same spin and isospin such as $\omega$ and $\rho$ and $\eta$ and $\eta'$ we further assume that the overlaps, say $< n| vac,\,a\,\bar b\, attached\,>$ are related by the mixing angles.

We need as said to be careful in treating eq. (\ref{17}) since the matrix elements

\be
\nonumber<vac, \bar c\,d\,removed\,| vac,\,a\,\bar b\, attached\,>
\ee
 in general are diverging. For some of them we can, however, determine their ratios to each other. To do this we use the self-consistency eq. (\ref{16}) in the limit $R _1 \rightarrow \infty$ and $R_2 \rightarrow 0$. When $R _1 \rightarrow \infty$ the exponentials in the righthand side tend to zero unless there is a state $|n>$ with mass zero. We will hence work in a limit of the underlying hadron model where the pion is massless.

Hence if there is a pion coupling in the crossed channel we obtain
\begin{align}
\nonumber
&< vac, \bar c\,d\,removed\,| vac,\,a\,\bar b\, attached\,> \\
& \longrightarrow_{R_1 \rightarrow \infty, R_2 \rightarrow 0} -\frac{\tau'(R_1) R_1}{ \tau'(R_2) R_2} \displaystyle \sum_{\pi^+,\pi^0,\pi^-} < vac, \bar b\,d\,removed\,| \pi>\,< \pi | vac,\,a\,\bar c\, attached\,>.
\end{align}

By using Clebsch-Gordan coefficients we could then compute the following ratios
\begin{align}\label{19}
\nonumber &< vac, \sqrt{1/2}[\bar u\downarrow d\uparrow - \bar u\uparrow d\downarrow] removed | vac, \sqrt{1/2}[ u\uparrow \bar d\downarrow - \ u\downarrow \bar d\uparrow] attached>:\\
\nonumber&< vac, \sqrt{1/2}[\bar u\uparrow d\downarrow + \bar u\uparrow d\downarrow] removed | vac, \sqrt{1/2}[ u\uparrow \bar d\downarrow + \ u\downarrow \bar d\uparrow] attached>:\\
\nonumber &< vac, \sqrt{1/2}[\bar u\uparrow u\uparrow - \bar d\uparrow d\uparrow] removed | vac, \sqrt{1/2}[ u\downarrow \bar u\downarrow - \ d\downarrow \bar d\downarrow] attached>:\\
\nonumber &< vac, \sqrt{1/4}[\bar u\downarrow u\uparrow + \bar d\downarrow d\uparrow-\bar u\uparrow u\downarrow - \bar d\uparrow d\downarrow] removed|\\
\nonumber &|vac, \sqrt{1/4}[\bar u\uparrow u\downarrow +  d\uparrow \bar d\downarrow- u\downarrow \bar u\uparrow -   d\downarrow \bar d\uparrow]  attached>:\\
\nonumber &< vac, \sqrt{1/2}[\bar u\downarrow u\downarrow + \bar d\downarrow d\downarrow] removed | vac, \sqrt{1/2}[ u\uparrow \bar u\uparrow +\ d\uparrow \bar d\uparrow] attached>:\\
&=1:1:(-1):(-3):3.
\end{align}

The choices of quarks (\ref{19}) are such that the combinations have the quantum numbers of $\pi$, $\rho^+(S_z = 0)$, $\rho^ 0(S_z = -1)$, $\eta_{ud}$ and $\omega (S_z = 1)$ respectively. Here we used the notation
\be
\eta_{ud}=\sqrt{\frac{1}{3}}\eta + \sqrt{\frac{2}{3}}{\eta'}.
\ee
The calculation is a classical quark model calculation where we have here used the following conventions. The isodoublet of the antiquarks is $(-\bar d, \bar u)$. We have chosen the phases such that $<vac, J^1, J^1_z, I^1, I^1_z \,removed| vac, J^2, J^2_z, I^2, I^2_z\, attached>$ is proportional to the Clebsh-Gordan coefficients  $<0,0| J^1, J^1_z, I^1, I^1_z > <0,0| J^2, J^2_z, I^2, I^2_z>$. The ket vectors $|J, J_z) $ and $|I, I_z) $ are defined by using the Condon-Shortley phase conventions taking the quarks in the order written.

\subsection{Comparison to the real world}
In principle we can now use the result above and look for modular functions to solve for (\ref{16}) in the various channels for which the quark quantum numbers give mesonic states. Even some 45 years ago there were tentalizing experimental facts in hadron physics which pointed to a string-like behaviour of the mesons, above all the nearly linearly rising Regge trajectories. Even if there were a hadronic string theory for mesons we know that it has to be unitarized and then there will be mass corrections, not only imaginary parts.

Another fact is the (almost) quantised masses of the mesons built by the $u$ and $d$ quarks. To a very good approximations the masses are (remember $\alpha'$ is set to $1$).

\begin{align}\label{21}
\nonumber &m_\pi^2 =0\\
\nonumber &m_\eta^2 = 1/4\\
\nonumber &m_\rho^2 = 1/2\\
\nonumber &m_\omega^2= 1/2\\
 &m_{\eta'}^2 = 3/4.
 \end{align}
With these numbers introduced in (\ref{16}) we could search the literature for possible modular functions. Our somewhat amateurish searches did not find any such functions. 

There is though one prediction that we can make directly. Since we assume that the underlying string theory is critical in $d=4$ we know that in critical string theories the first daughter trajectory is missing, because the norms of the states on it are proportional to $d-d_{crit}$. Hence there are no spin-$0$ state with the same mass as the $\rho$ and $\omega$.

However there was one test that we could do in this case. We could consider the case when the two sides of the rectangle in Fig. 3 are the same. Since the horizontal and vertical channels now carry different mesons this is a non-trivial test. In the case $R_1=R_2=\pi$ we get the following relation

\begin{align}
\nonumber&-\frac{1}{2} < vac, \sqrt{\frac{1}{2}}(\bar u \downarrow d \uparrow-\bar u\uparrow d\downarrow)\, removed|\pi^+>E^{-m_\pi^2 \pi} <\pi^+|vac , \sqrt{\frac{1}{2}}(u\uparrow \bar d\downarrow-u\downarrow\bar d \uparrow) \, attached>\\
\nonumber&+\frac{1}{2} < vac, \sqrt{\frac{1}{2}}(\bar u \downarrow d \uparrow+\bar u \uparrow d\downarrow)\, removed|\rho^+(S_z=0)>E^{-m_\rho^2 \pi} \\
\nonumber&<\rho^+(S_z=0)|vac , \sqrt{\frac{1}{2}}(u\uparrow \bar d\downarrow+ u\downarrow \bar d\uparrow)\, attached>+...\\
\nonumber&= -\frac{1}{2} < vac, \sqrt{\frac{1}{2}}(\bar d \uparrow d \uparrow-\bar u\uparrow u \uparrow) \, removed|\omega(S_z=-1)>E^{-m_\omega^2 \pi} \\
\nonumber &<\omega(S_z=1)|vac ,\sqrt{\frac{1}{2}} (u\downarrow \bar u\downarrow+d\downarrow \bar d\downarrow)\, attached>\\
\nonumber & +\frac{1}{2} < vac, \sqrt{\frac{1}{2}}(\bar u \uparrow u \uparrow-\bar d \uparrow d \uparrow) \, removed|\rho^0(S_z=-1)>E^{-m_\rho^2 \pi}\\
& <\rho^0(S_z=-1)|vac , \sqrt{\frac{1}{2}}(u\downarrow \bar u\downarrow-d\downarrow \bar d \downarrow)\, attached>\ +..
\end{align}

By manipulating this equations with the calculations above we arrive at

\be
\nonumber-\frac{1}{2} e^{-m_\pi^2 \pi} + \frac{1}{2} e^{-m_\rho^2 \pi} +..= -\frac{3}{2}e^{-m_\omega^2 \pi} +\frac{1}{2} e^{-m_\rho^2 \pi}+..
\ee
or
\be
-\frac{1}{2} e^{-m_\pi^2 \pi}+e^{-m_\rho^2 \pi}+\frac{3}{2}e^{-m_\omega^2 \pi}\approx 0.
\ee
There is one more combination of the up- and downquarks that gives a non-trivial relation

\be
e^{-m_\rho^2 \pi} - \frac{1}{3} e^{-m_\eta^2 \pi} -\frac{2}{3} e^{-{m_\eta'}^2 \pi}\approx 0.
\ee
These relations are correct to within a few percent when we use the masses (\ref {21}), and slightly worse if we use the actual masses, which we did in the paper~\cite{Brink:1974cm}. The correct thing in a hypothesised string theory is to use  the masses (\ref {21}) though. A partition function must have quantised coefficients in the exponentials. 

In principle we could form full partition functions using the Regge trajectories up to infinite masses. They will not be modular functions but could still have interesting mathematical properties. The corrections from the higher particles on the trajectories to our calculations are very small. One interesting aspects of the calculations is that both the coefficients in the series which come from the quark model and hence group theory, and the exponentials which come from the string assumptions must cooperate to get formulae that work. This is reminiscent of moonshine. We can also phrase it to say that the string assumption forces the masses of the mesons made up with up and down quarks to be quantised as in (\ref {21}). (We should remember though that we got two relations for five masses.)

What happens when we put strange quarks at the ends? One problem to start with is that the strange mesons do not seem to have masses so close to quantised numbers. We can now understand that from the strange quark having a mass which cannot be neglected. However, also the strange mesons have quite linear Regge trajectories indicating also some string-like behaviour. We never attempted to set up corresponding partition functions involving strange quarks. (There were ideas that the bare kaon mass should be $1/4$ and that a string model involving strangeness could be like a square root of the RNS-model. None was of course found, but it is our duty to speculate.) 

In a sense we might already have learnt as much as we can from our assumptions about strings with quarks at the endpoints. 

\subsection{Further results from the string assumption}
Since the string assumption works so well in certain cases we can remind ourselves of other examples. 

Already in the infancy of dual models/string theory Lovelace~\cite{Lovelace:1969se} and Joel Shapiro~\cite{Shapiro:1969km} found an amplitude for pion-pion scattering in the form

\be
A(s, t) = g^2 \,\,\frac{\Gamma(1-\alpha(s)) \Gamma(1-\alpha(t))}{\Gamma(1-\alpha(s) -\alpha(t))},
\ee
where $\alpha(s) = \frac{1}{2}+\alpha' s$ is the $\rho$-trajectory. We take the pion mass to be zero.

This was a seemingly good amplitude with the correct intercept for the trajectory, but this was not the amplitude that could be extended to $N$-point amplitudes. (It is the amplitude in the Neveu-Schwarz model when the intercept is $1$.) For a long time it was a solitary amplitude in search of a family. However, it has been found that it has very interesting properties, namely that it has critical dimension $4$ and is seemingly ghost-free~\cite{PDV}, \cite{Veneziano:2017cks}. This is in line with our result above where the $\pi$ and $\rho$ are well described by strings with quarks at the endpoints. This picture will break down somewhere which might be an explanation why we cannot make it into a fully unitarizable model.

Coming back to the original Veneziano amplitude~\cite{Veneziano:1968yb} for the scattering amplitude $\pi\,\,\pi \rightarrow \pi\,\,\omega$
\be
A(s, t) =\epsilon_{\mu \nu \sigma\tau}\,\, e^\mu p_1^\nu p_2^\sigma p_3^\tau\,\, g^2 \,\,\Gamma(1-\alpha(t)) (-\alpha(s))^{\alpha(t)-1},
\ee
it has also been conjectured to have critical dimension four and to be ghost free. This is also in line with the arguments above.

\subsection{Further attempts to introduce strings with flavour symmetry}

The Ramond-Neveu-Schwarz model introduced supersymmetry on the world sheet. When supersymmetry was introduced also for $d=4$ field theories it became clear that supersymmetry could carry an internal symmetry. In the string picture that would mean that the flavour quantum number was distributed along the string, quite opposite to the picture I gave above. This does not sound now to be a very interesting option, but it opened up for the possibility to construct new string models. Together with Paolo Di Vecchia and a number of collaborators we followed this thread and managed to first construct new Virasoro algebras~\cite{Ademollo:1975an} and found two possible new models, one based on an $U(1)$ internal symmetry~\cite{Ademollo:1976pp} and one with an $SU(2)$ internal symmetry~\cite{Ademollo:1976wv}. 

The first one described a string living in a four-dimensional space with two time coordinates and the second had ghosts in any dimensions. The virtue of this exercise is that it showed the possible strings models there are. It did strengthen our belief that the truly unitarizable string theories should instead be used for unifying theories involving gravity. Interpreted this way all the problems in the Ramond-Neveu-Schwarz model became virtues and the rest is history.

\section{A retrospective view of the results}

Hadronic string theories became quickly obsolete when QCD was discovered~\cite{Gross:1973id}. Finally we had a theory for the strong interactions. To describe hadron physics we can use lattice gauge theories~\cite{Wilson:1974sk},~\cite{Creutz:1980zw} and/or chiral perturbation theory~\cite{Weinberg:1978kz}. 

In both approaches it is important to have non-zero quark masses, and the pion mass  which is small due the spontaneous symmetry breaking of the chiral symmetry has to be taken to be non-zero. In QCD there is a problem with the $U(1)$ anomaly leading to the $\eta'$ mass. It would be hard to find quantised masses in those approaches. In our approach we see the problem from a different angle representing the quark-antiquark interactions, ie. the underlying QCD, by a relativistic string, and that points towards (approximately) quantised masses for the mesons built up by the $u$ and $d$ quarks in $SU(3)$. Those masses are quite unnatural in lattice QCD even though Wilson in his original paper described mesons as a flux tube with quarks at the ends.

It should be pointed out also that a fully unitary model with a massless scalar would be very different from present day hadron physics. This is yet another reason why we should not expect to be able to fully unitarize our string model.

The flux tube, ie. the closed string in Fig 1, became of course the graviton trajectory in the Superstring Theory. For a unitarizable hadronic theory it should be the pomeron trajectory. It should have intercept $1$ instead of $2$ for the Superstring. There is a four-point amplitude first constructed by Virasoro~\cite{Virasoro:1969me} and Shapiro~\cite{Shapiro:1969km} that could be seen as a candidate for such an elastic scattering. It is not known if such an amplitude could be critical in $4$ dimensions of spacetime. In perturbative QCD the pomeron is very different from this picture and is described by gluon exchanges or glue balls. The fact that the Lovelace-Shapiro amplitude is seemingly not unitarizable can be seen as a sign that the elastic scattering in hadron physics cannot be described by a pomeron exchange with a trajectory of real particles. 

The quantised masses point towards some extra symmetry. Note that the mass formulae is quite different from the Gell-Mann-Okubo mass formula~\cite{MGM},~\cite{Okubo:1961jc}. That formula is linear, and with three free parameters it relates the masses in the octets and decouplets. It is much more related to the quark structure and quark masses.

We have a theory for the strong interactions, QCD. It has been enormously successful. However, the strong interactions is  one of the most complicated dynamical systems that we have tried to solve in all details.  Not all of those are accessible straightforwardly from the pure theory. The hadronic string picture is approximate but in some circumstances a very useful picture that gives new insight in hadron physics. This is what I have wanted to show in this paper.


\begin{thebibliography}{99}
\bibitem{Freund:1971tw} 
  P.~G.~O.~Freund, H.~F.~Jones and R.~J.~Rivers,
 ``Dynamics versus selection rules in diffraction dissociation,''
  Phys.\ Lett.\  {\bf 36B}, 89 (1971).
  \bibitem{Freund:1972gw} 
  P.~G.~O.~Freund,
 ``Quark spin in a dual-resonance model,''
  Nuovo Cim.\ A {\bf 8}, 525 (1972).
  \bibitem{Freund:1980xh} 
  P.~G.~O.~Freund and M.~A.~Rubin,
  ``Dynamics of Dimensional Reduction,''
  Phys.\ Lett.\  {\bf 97B}, 233 (1980).
  \bibitem{Freund:1987ks} 
  P.~G.~O.~Freund and M.~Olson,
  ``p-ADIC DYNAMICAL SYSTEMS,''
  Nucl.\ Phys.\ B {\bf 297}, 86 (1988).
  \bibitem{Freund:1987ck} 
  P.~G.~O.~Freund and E.~Witten,
 ``Adelic String Amplitudes,''
  Phys.\ Lett.\ B {\bf 199}, 191 (1987).
  \bibitem{Veneziano:1968yb} 
  G.~Veneziano,
  ``Construction of a crossing - symmetric, Regge behaved amplitude for linearly rising trajectories,''
  Nuovo Cim.\ A {\bf 57}, 190 (1968).
  \bibitem{Chan:1969ex} 
  H.~M.~Chan and S.~T.~Tsou,
  ``Explicit construction of the n-point function in the generalized Veneziano model,''
  Phys.\ Lett.\ B {\bf 28}, 485 (1969).
  \bibitem{Koba:1969rw} 
  Z.~Koba and H.~B.~Nielsen,
 ``Reaction amplitude for n mesons: A Generalization of the Veneziano-Bardakci-Ruegg-Virasoro model,''
  Nucl.\ Phys.\ B {\bf 10}, 633 (1969).
   \bibitem{Nambu:1969se} 
  Y.~Nambu,
  ``Quark model and the factorization of the Veneziano amplitude,''
  In *Detroit 1969, Proceedings, Conference On Symmetries.
   \bibitem{Susskind:1970xm} 
  L.~Susskind,
 ``Dual symmetric theory of hadrons. 1.,''
  Nuovo Cim.\ A {\bf 69}, 457 (1970).
  \bibitem{holger} H.~B.~Nielsen, XV Int. Conf.on High Energy Physics, Kiev (1970).
 \bibitem{Virasoro:1969zu} 
  M.~S.~Virasoro,
 ``Subsidiary conditions and ghosts in dual resonance models,''
  Phys.\ Rev.\ D {\bf 1}, 2933 (1970).
 \bibitem{Fubini:1969wp} 
  S.~Fubini, D.~Gordon and G.~Veneziano,
 ``A general treatment of factorization in dual resonance models,''
  Phys.\ Lett.\ B {\bf 29}, 679 (1969).
\bibitem{Ramond:1971gb} 
  P.~Ramond,
 ``Dual Theory for Free Fermions,''
  Phys.\ Rev.\ D {\bf 3}, 2415 (1971).
  \bibitem{Neveu:1971rx} 
  A.~Neveu and J.~H.~Schwarz,
``Factorizable dual model of pions,''
  Nucl.\ Phys.\ B {\bf 31}, 86 (1971).
  \bibitem{Thorn:1971jc} 
  C.~B.~Thorn,
 ``Embryonic Dual Model for Pions and Fermions,''
  Phys.\ Rev.\ D {\bf 4}, 1112 (1971).
  \bibitem{Corrigan:1972tg} 
  E.~Corrigan and D.~I.~Olive,
"Fermion meson vertices in dual theories,''
  Nuovo Cim.\ A {\bf 11}, 749 (1972).
  \bibitem{Brink:1973jd}
   L.~Brink, D.~I.~Olive, C.~Rebbi and J.~Scherk,
 ``The Missing Gauge Conditions for the Dual Fermion Emission Vertex and Their Consequences,''
  Phys.\ Lett.\  {\bf 45B}, 379 (1973).
 \bibitem{Brink:1973qm} 
  L.~Brink and D.~I.~Olive,
 ``The physical state projection operator in dual resonance models for the critical dimension of space-time,''
  Nucl.\ Phys.\ B {\bf 56}, 253 (1973).
 \bibitem{Gliozzi:1976jf} 
  F.~Gliozzi, J.~Scherk and D.~I.~Olive,
 ``Supergravity and the Spinor Dual Model,''
  Phys.\ Lett.\  {\bf 65B}, 282 (1976).
 \bibitem{Lovelace:1971fa} 
  C.~Lovelace,
``Pomeron form-factors and dual Regge cuts,''
  Phys.\ Lett.\  {\bf 34B}, 500 (1971).
  \bibitem{Brower:1972wj} 
  R.~C.~Brower,
 ``Spectrum generating algebra and no ghost theorem for the dual model,''
  Phys.\ Rev.\ D {\bf 6}, 1655 (1972).
\bibitem{Goddard:1972ky} 
  P.~Goddard, C.~Rebbi and C.~B.~Thorn,
 ``Lorentz covariance and the physical states in dual resonance models,''
  Nuovo Cim.\ A {\bf 12}, 425 (1972).
  \bibitem{Goddard:1973qh} 
  P.~Goddard, J.~Goldstone, C.~Rebbi and C.~B.~Thorn,
 ``Quantum dynamics of a massless relativistic string,''
  Nucl.\ Phys.\ B {\bf 56}, 109 (1973).
  \bibitem{NG}
  Y.~Nambu, "Duality and Hadrodynamics," Notes prepared for the Copenhagen High Energy Symposium, Aug. 1970.
  \bibitem{Goto:1971ce} 
   T.~Got\={o},
  ``Relativistic quantum mechanics of one-dimensional mechanical continuum and subsidiary condition of dual resonance model,''
  Prog.\ Theor.\ Phys.\  {\bf 46}, 1560 (1971).
  \bibitem{Brink:1973gi} 
  L.~Brink and D.~I.~Olive,
 ``Recalculation of the the unitary single planar dual loop in the critical dimension of space time,''
  Nucl.\ Phys.\ B {\bf 58}, 237 (1973).
  \bibitem{Fairlie:1970tc} 
  D.~B.~Fairlie and H.~B.~Nielsen,
 ``An analogue model for ksv theory,''
  Nucl.\ Phys.\ B {\bf 20}, 637 (1970).
  \bibitem{Nielsen:1969} 
  H.~B.~Nielsen,
  ``An Almost Physical Interpretation of the n Point Veneziano Amplitude,''
  NORDITA preprint (1969).
  \bibitem{Brink:1973kn} 
  L.~Brink and H.~B.~Nielsen,
 ``A physical interpretation of the jacobi imaginary transformation and the critical dimension in dual models,''
  Phys.\ Lett.\  {\bf 43B}, 319 (1973).
  \bibitem{DelGiudice:1971yjh} 
  E.~Del Giudice, P.~Di Vecchia and S.~Fubini,
 ``General properties of the dual resonance model,''
  Annals Phys.\  {\bf 70}, 378 (1972).
  \bibitem{Misha}
  L.~Brink, N.~Wyllard and M.~A.~Vasiliev, unpublished (1997).
  \bibitem{HBNLB2}
  L.~Brink and H.~B.~Nielsen, unpublished (1972).
  \bibitem{Nahm:1976dj} 
  W.~Nahm,
 ``Mass Spectra of Dual Strings,''
  Nucl.\ Phys.\ B {\bf 114}, 174 (1976).
  \bibitem{Gross:1984dd} 
  D.~J.~Gross, J.~A.~Harvey, E.~J.~Martinec and R.~Rohm,
 ``The Heterotic String,''
  Phys.\ Rev.\ Lett.\  {\bf 54}, 502 (1985). \bibitem{Brink:1974cm} 
  L.~Brink and H.~B.~Nielsen,
 ``Two Mass Relations for Mesons from String - Quark Duality,''
  Nucl.\ Phys.\ B {\bf 89}, 118 (1975).
  \bibitem{Lovelace:1969se} 
  C.~Lovelace,
  ``A novel application of regge trajectories,''
  Phys.\ Lett.\  {\bf 28B}, 264 (1968).
  \bibitem{Shapiro:1969km} 
  J.~A.~Shapiro,
 ``Narrow-resonance model with regge behavior for pi pi scattering,''
  Phys.\ Rev.\  {\bf 179}, 1345 (1969).
   \bibitem{PDV}
  A. Cappelli et al, "The birth of string theory", Cambridge University Press, p. 585 (2012). 
  \bibitem{Veneziano:2017cks} 
  G.~Veneziano, S.~Yankielowicz and E.~Onofri,
  ``A model for pion-pion scattering in large-N QCD,''
  JHEP {\bf 1704}, 151 (2017).
  \bibitem{Ademollo:1975an} 
  M.~Ademollo {\it et al.},
  ``Supersymmetric Strings and Color Confinement,''
  Phys.\ Lett.\  {\bf 62B}, 105 (1976).
  \bibitem{Ademollo:1976pp} 
  M.~Ademollo {\it et al.},
 ``Dual String with U(1) Color Symmetry,''
  Nucl.\ Phys.\ B {\bf 111}, 77 (1976).
  \bibitem{Ademollo:1976wv} 
  M.~Ademollo {\it et al.},
 ``Dual String Models with Nonabelian Color and Flavor Symmetries,''
  Nucl.\ Phys.\ B {\bf 114}, 297 (1976).
  \bibitem{Gross:1973id} 
  D.~J.~Gross and F.~Wilczek,
 ``Ultraviolet Behavior of Nonabelian Gauge Theories,''
  Phys.\ Rev.\ Lett.\  {\bf 30}, 1343 (1973).
  \bibitem{Wilson:1974sk} 
  K.~G.~Wilson,
  ``Confinement of Quarks,''
  Phys.\ Rev.\ D {\bf 10}, 2445 (1974).
  \bibitem{Creutz:1980zw} 
  M.~Creutz,
 ``Monte Carlo Study of Quantized SU(2) Gauge Theory,''
  Phys.\ Rev.\ D {\bf 21}, 2308 (1980).
  doi:10.1103/PhysRevD.21.2308.
  \bibitem{Weinberg:1978kz} 
  S.~Weinberg,
 ``Phenomenological Lagrangians,''
  Physica A {\bf 96}, no. 1-2, 327 (1979).
  \bibitem{Virasoro:1969me} 
  M.~A.~Virasoro,
 ``Alternative constructions of crossing-symmetric amplitudes with regge behavior,''
  Phys.\ Rev.\  {\bf 177}, 2309 (1969).
  \bibitem{MGM}
.M:~Gell-Mann, "The Eightfold Way: A Theory of Strong Interaction Symmetry". Synchrotron Laboratory Report CTSL-20. California Institute of Technology (1961).
\bibitem{Okubo:1961jc} 
  S.~Okubo,
 `Note on unitary symmetry in strong interactions,''
  Prog.\ Theor.\ Phys.\  {\bf 27}, 949 (1962).
  doi:10.1143/PTP.27.949.


 

\end{thebibliography}
\end{document}